\documentclass[%
reprint,
 amsmath,amssymb,
 aps,pre
]{revtex4-2}

\usepackage{graphicx}
\usepackage{dcolumn}
\usepackage{bm}
\usepackage{hyperref}
\usepackage[mathlines]{lineno}
\usepackage{xcolor}

\begin{document}

\preprint{APS/123-QED}

\title{Topological defects in spiral wave chimera states}

\author{Lintao Liu}
\author{Nariya Uchida}%
 \email{uchida@cmpt.phys.tohoku.ac.jp}
\affiliation{%
 Department of Physics, Tohoku University, Sendai 980-8578, Japan}

\date{\today}

\begin{abstract}
Chimera states, characterized by the coexistence of coherent and incoherent domains, represent a paradigm of self-organization in complex systems. In this study, we introduce a topological analysis method based on winding numbers to characterize the dynamics of spiral wave chimeras in a two-dimensional phase oscillator network. Our investigation reveals distinct scaling laws governing the system's evolution across the phase lag $\alpha$. Perturbation analysis in the limit $\alpha \to 0$ demonstrates that the incoherent core radius scales linearly with $\alpha$. In contrast, within the stable chimera regime, the average total positive winding number $\mu$ follows a clear exponential growth law $\mu = ae^{b\alpha}$. 
This scaling disparity signals a physical crossover from a regime dominated by geometric core expansion to one driven by active topological excitation. Furthermore, we identify a statistical transition in the defect distribution from binomial-like to Poisson-like behavior at a critical threshold $\alpha^*$. These results demonstrate that topological defects possess intrinsic statistical order, establishing $\mu$ as a robust macro-variable for analyzing the structural complexity of chimera states.
\end{abstract}

\maketitle


{\it Introduction.}
Large systems of coupled oscillators are a fundamental model for understanding cooperative behaviors in a wide range of physical, chemical, and biological systems~\cite{Kuramoto1984,strogatz2003sync}. When the coupling is sufficiently weak, the complex dynamics can be simplified and described by a phase variable through phase reduction. Depending on the form of the coupling function, these oscillators exhibit not only global synchronization and desynchronization~\cite{RevModPhys.77.137}, but also chimera states~\cite{kuramoto2002coexistence,PhysRevLett.93.174102}, which are characterized by the spatial coexistence of coherent and incoherent regions. Initially found in non-locally coupled one-dimensional systems, chimera states have since been observed in two- and three-dimensional systems~\cite{PhysRevE.69.036213,PhysRevE.70.065201,article23,article33,PhysRevE.94.010204}, as well as in populations of oscillators~\cite{PhysRevLett.101.084103}. They have been experimentally realized in chemical~\cite{PhysRevLett.110.244102,Totz2018,Tinsley2012}, mechanical~\cite{doi:10.1073/pnas.1302880110,Kapitaniak2014}, and electronic oscillators~\cite{PhysRevE.90.030902,PhysRevE.90.032905} and are particularly relevant to the study of neuronal networks~\cite{MAJHI2019100}.

To investigate these phenomena, we employ the non-local Sakaguchi-Kuramoto model~\cite{10.1143/PTP.76.576} on a two-dimensional \(N \times N\) square lattice, 
where each lattice point hosts an identical oscillator. 
Each oscillator interacts with \(N(R)\) neighboring oscillators within a finite distance \(R\).
The governing equation is given by:
\begin{equation}
    \frac{d \phi(\bm{r}) }{dt} = \omega - \frac{K}{N(R)} 
    \sum_{\bm{r}', R} 
    \sin\left[\phi(\bm{r}) - \phi(\bm{r}')  + \alpha\right],
    \label{eq.R}
\end{equation}
where $\phi(\bm{r})$ denotes the phase of the oscillator at position $\bm{r}$, 
$\omega$ is the natural frequency which we assume constant, $K$ is the coupling strength,
\(\alpha\) is the phase lag,
and $\sum_{\bm{r}', R}$ means that the sum is taken over the neighbors $\bm{r}'$
in the range $0 < |\bm{r} - \bm{r}'| \leq R$.
We simplify the equation by applying the transformation \(\phi' = \phi - \omega t\) 
and \(\tau = t / K\), as
\begin{equation}
    \frac{d \phi(\bm{r}) }{d\tau} = - \frac{1}{N(R)} 
    \sum_{\bm{r}', R} 
    \sin\left[\phi(\bm{r}) - \phi(\bm{r}')  + \alpha\right].
    \label{eq.RR}
\end{equation}
Since the equation remains invariant under the transformation \(\alpha \to -\alpha\) 
and \(\phi \to -\phi\), we restrict the phase lag 
parameter \(\alpha\) to the interval \([0, \pi)\) without loss of generality.
We define the local complex order parameter to characterize the system's dynamics as
\begin{equation}
    z(\bm{r}) = |z|(\bm{r}) e^{i \Phi(\bm{r})} = \frac{1}{N(R)+1} \left|\sum_{\bm{r}',R} 
    \exp [i\phi(\bm{r}')] \right|.
\end{equation}

In two dimensions, a rich variety of patterns can emerge, among which spiral wave chimeras are the most archetypal example~\cite{PhysRevE.70.065201,PhysRevLett.104.044101} and have been observed experimentally~\cite{totz2018spiral}.
The Kuramoto-like model on a two-dimensional lattice shares a powerful analogy 
with the XY model of planar spins. The study of two-dimensional Kuramoto-like models 
from a vortex perspective has recently gained traction, with several studies 
investigating vortex dynamics. These works have explored the Kuramoto-like model 
for specific cases, such as nearest-neighbor coupling ($R=1$) with random intrinsic 
frequencies and the entrainment transition~\cite{PhysRevE.82.036202}, 
or through a duality transformation to expose underlying vortex structures~\cite{PhysRevE.103.032204} 
and in 2D Spin Torque Oscillators~\cite{Flovik_2016}. 
Additionally, research on the 2D-XY model of self-driven rotors~\cite{PhysRevLett.127.088004} has provided insights into defect dynamics and their superdiffusive behavior. 
Furthermore, the dynamics of traveling spiral wave chimeras and their phase-lag dependent transitions have also been studied in detail~\cite{bataille2023traveling}.
However, to the best of our knowledge, the specific case of a non-local coupling ($R>1$) with a non-zero phase lag has not been studied from a topological perspective. This particular combination of parameters is crucial for the emergence of chimera states. The primary objective of this paper is to investigate these chimera states through a topological lens.

In this paper, we study the topological defects associated with the spiral wave chimera states.
In Sec. II, we will first present analytical results for a small-$\alpha$ regime.
Sec. III provides numerical simulation results, focusing on the emergence of
spatiotemporal patterns and then on the formation of topological defects.
In Sec. IV, we analyze the statistical properties of the defects and show that 
the average total positive winding number exhibits an exponential growth 
with respect to the phase lag $\alpha$ and propose an entropy hypothesis to explain the system's properties.

{\it Analytical Results.}
Following Ref.~\cite{PhysRevE.69.036213}, we rewrite 
the governing equation (\ref{eq.RR}) in the form of self-consistent equation
\begin{eqnarray}
|z|(\bm{r})e^{i \Phi(\bm{r})} &=& ie^{-i \alpha} \int_{\mathbb{R}^2} G(|\bm{r}- \bm{r}'|) \exp[i \Phi(\bm{r}')]  \nonumber \\
 & &\times \frac{\Delta - \sqrt{\Delta^2 - |z|(\bm{r}')^2}}{|z|(\bm{r}')} d\bm{r}'.
\end{eqnarray}
where $\Delta = \omega - \Omega$, with the kernel function given by  
\begin{equation}
    G(|\boldsymbol{x}- \boldsymbol{x}'|) = 
    \begin{cases}
        \frac{1}{\pi R^2}, & \text{if } |\boldsymbol{x}- \boldsymbol{x}'| < R, \\
        0, & \text{if } |\boldsymbol{x}- \boldsymbol{x}'| > R.
    \end{cases}
\end{equation}
Following the approach in Ref.~\cite{PhysRevLett.104.044101}, we adopt the Spiral Wave Ansatz~\cite{15cd7884-e7e8-3592-aee4-ba066c0c3e21}:  
\begin{equation}
    |z|(\boldsymbol{x}) = A(r), \quad \Phi(\boldsymbol{x}) = \theta + \Psi(r).
\end{equation}
Substituting this ansatz, we obtain:
\begin{eqnarray}
A(r)e^{i \Psi(r)} &=& ie^{-i \alpha} \int_0^\infty \exp[i \Psi(s)]   \nonumber \\
 & &\times \frac{\Delta - \sqrt{\Delta^2 - A^2(s)}}{A(s)} 
    K(r,s) \, ds,
\end{eqnarray}
where the integral kernel \( K(r,s) \) is given by  
\begin{equation}
    K(r,s) = 2s \int_0^\pi G(\sqrt{r^2+s^2 - 2rs \cos \theta})
    \cos\theta \, d\theta.
\end{equation}
The self-consistent equations remain analytically intractable. 
To proceed, a series expansion in the small parameter $\alpha$ is needed, following the method in Ref.~\cite{PhysRevLett.104.044101}:
\begin{equation}
    \begin{aligned}
        \Delta &= \Delta_1 \alpha + O(\alpha^2), \\
        A(r) &= A_0(r) + A_1(r) \alpha +  O(\alpha^2), \\
        \Psi(r) &= \Psi_0(r) + \Psi_1(r)\alpha + O(\alpha^2).
    \end{aligned}
\end{equation}

At leading order $\mathcal{O}(1)$, the self-consistency equation reduces to
\begin{equation}
\begin{aligned}
A_0(r)  &= \int_0^\infty K(r,s) \, ds,\\
A_0(r) \Psi_0(r) &= \int_0^\infty K(r,s) \Psi_0(s) \, ds.
\end{aligned}
\end{equation}
By rotational symmetry of the system, we choose $\Psi_0(r) = 0$ without loss of generality. 
Substituting the expression for the kernel and applying the variable transformation $u = s^2$, we obtain the following expression for $A_0(r)$:
\begin{equation}
A_0(r) =
\frac{1}{\pi R^2} \int_{(R-r)^2}^{(R+r)^2}
\sqrt{1 - \frac{(r^2 + u - R^2)^2}{4r^2 u}}  du.
\label{eq:A}
\end{equation}
As $r \to 0$, the integral domain shrinks, $\lim_{r \to 0} A_0(r) = 0.$.
At $ r = R $, the integration range becomes $ 0 \leq u \leq 4R^2 $. Substituting into the integrand yields $ A_0(R) = \frac{8}{3\pi}$.
In the limit $r \gg R $, $\lim_{r \to \infty} A_0(r) = 1$ (see Supplementary Information S1 for details).

At $\mathcal{O}(\alpha)$ order, after simplification, we obtain $A_1(r) = 0$, and the equation becomes:
\begin{equation}
f(r) - \int_0^\infty \frac{K(r,s)}{A_0(s)} f(s)  ds 
= \Delta_1 \int_0^\infty \frac{K(r,s)}{A_0(s)}  ds - A_0(r).
\label{eq:f}
\end{equation}
where $f(r) = A_0(r) \Psi_1(r)$. This is a non-homogeneous Fredholm integral equation of the second kind, which generally requires numerical methods for solution. 

\begin{figure}[h]
\includegraphics[width=0.48\textwidth]{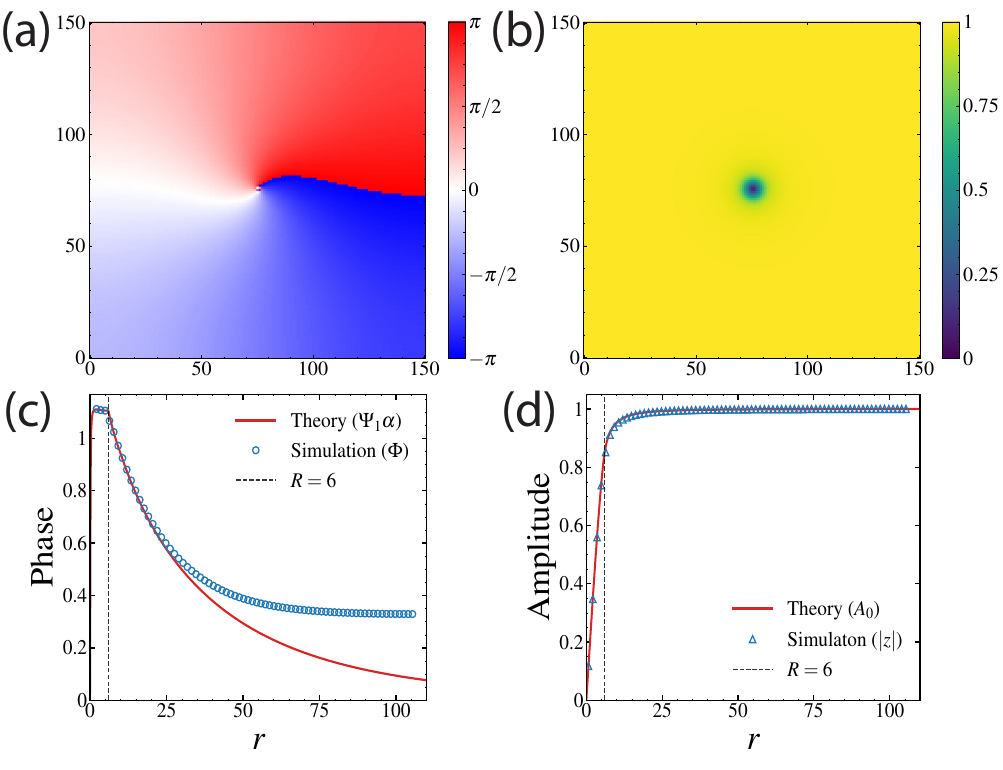}
\caption{Order parameter profiles.
Snapshots of (a) the phase $\Phi(\bm{r})$ and (b) amplitude $|z|(\bm{r})$
from the numerical simulation results for $N = 151$, $\alpha = 10^\circ$, and $t = 100$.
(c-d) Comparison between the analytical expressions Eqs.~(\ref{eq:A}),(\ref{eq:f}) and the diagonal cross-section of the numerical results [(a),(b)], where $r$ is 
the distance from the vortex.
\label{fig:OP}
}
\end{figure}

The incoherent oscillators are defined by the condition $A(r) < |\Delta|$. Let $r_{\text{core}}$ be the radius of the incoherent core. We approximate $A_0(r_{\text{core}})$ linearly as
$A_0(r_{\text{core}}) \approx A_0'(0) r_{\text{core}} = \Delta_1 \alpha$. Since $A_0(R) = \frac{8}{3\pi}$ and the function is approximately linear for small $r$, we estimate the derivative as $ A_0'(0) \approx \frac{8}{3\pi R}$. Substituting this into the previous expression and using $\Delta_1 \approx 1$, we find
\begin{equation}
r_{\text{core}} \approx \frac{1}{A_0'(0)} \alpha = \frac{3\pi}{8} R \alpha.
\label{eq:rho}
\end{equation}
This shows that the radius of the incoherent core grows approximately linearly with both $R$ and $\alpha$.

To evaluate the robustness of these patterns, we perform a linear stability analysis of the plane wave solutions~\cite{Li_2021,PhysRevE.70.065201}, defined by $\psi(\boldsymbol{r}, t) = \Omega t + \boldsymbol{Q} \cdot \boldsymbol{r}$, where $\boldsymbol{Q}$ denotes the wave vector and $\Omega$ represents the corresponding frequency (see Supplementary Information S2 for the full derivation). The local wave vector of the spiral arms is estimated via the phase gradient:$$\boldsymbol{Q} = \nabla \Phi = \frac{d \Psi(r)}{dr} \hat{\boldsymbol{r}} + \frac{1}{r} \hat{\boldsymbol{\theta}}$$Our analysis clarifies how the coupling range $R$ and the phase lag $\alpha$ synergistically dictate the stability of the spiral wave arms. In the in-phase regime ($0 < \alpha < \frac{\pi}{2}$), the stability domain is constrained to the approximate range $0 < \rho < 2.2917$, where $\rho = R|\boldsymbol{Q}|$. Conversely, in the anti-phase regime ($\frac{\pi}{2} < \alpha < \pi$), the stable domain shifts to $4.2005 < \rho < 6.1876$. Notably, as the coupling range $R$ increases, the stability criterion is more readily violated, which corresponds to a progressive shortening of the spiral arm wavelength. Consequently, within the regime $\frac{\pi}{2} < \alpha < \pi$, the spiral wave configurations often become intrinsically unstable, leading to the emergence of more complex spatiotemporal patterns.


\begin{figure*}
\includegraphics[width=0.95\textwidth]{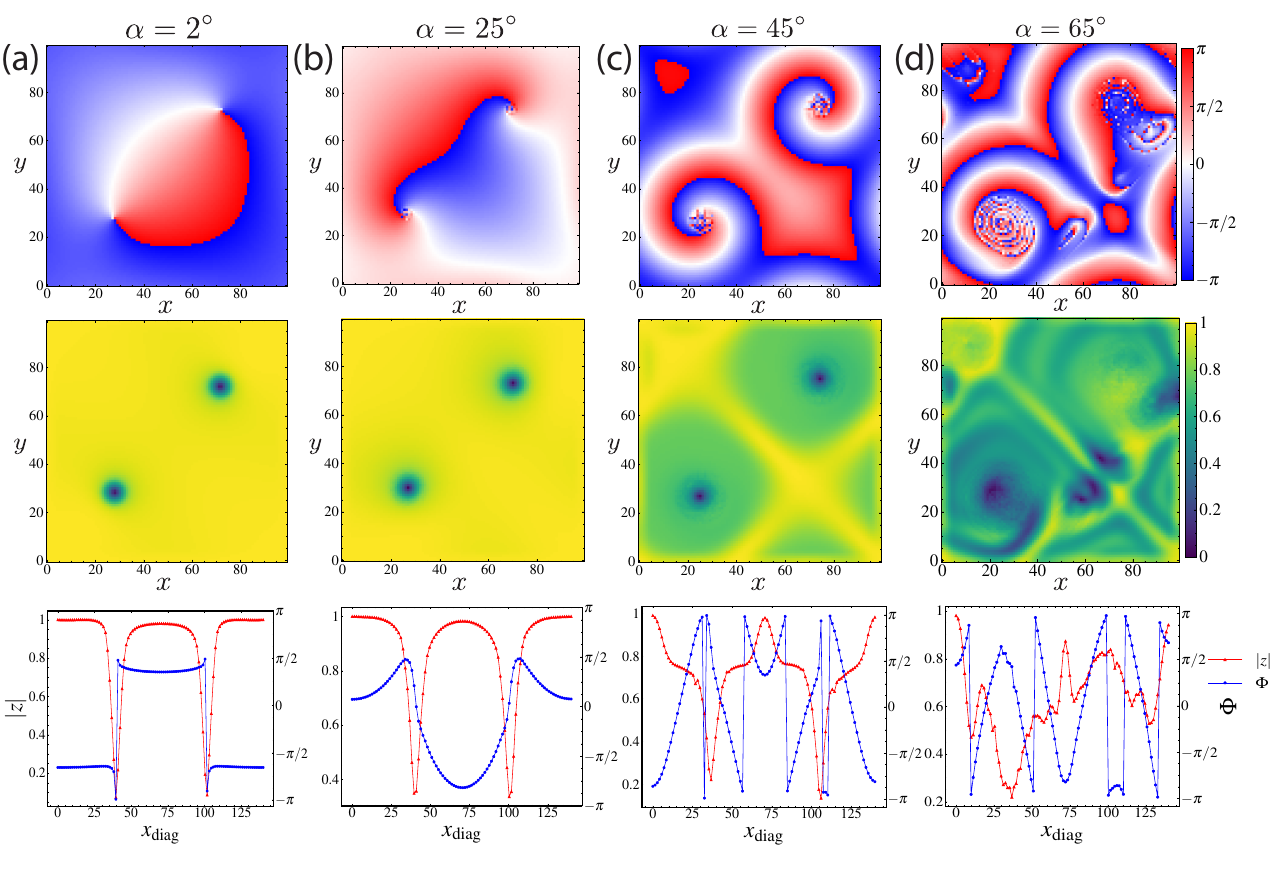}
\caption{\label{fig:2}
Spatial patterns of the phase $\phi(x,y)$ (first row), 
order parameter $|z|(x,y)$ (second row),
and the diagonal cross-section the order parameter profile sim$|z|(x_{\rm diag}, x_{\rm diag})$ (third row)
at \( t = 60 \) for different values of phase lag.  
(a) \( 0 < \alpha = 2^{\circ} < \alpha_0 \): A pair of spiral waves without a random core.  
(b) \( \alpha_0 < \alpha = 25^{\circ} < \alpha_1 \): A pair of spiral waves with an incoherent core.  
(c) \( \alpha_1 < \alpha = 45^{\circ} < \alpha_2 \): A pair of spiral waves with a stable incoherent core.  
(d) \( \alpha_2 < \alpha = 65^{\circ} < \pi/2 \): Irregular patterns and turbulence.
\label{fig:pairs}
}
\end{figure*}

{\it Numerical Simulation.}
We numerically simulate Eq.~(\ref{eq.RR}) using the fourth-order Runge-Kutta method for time integration with a time step of $dt = 0.02$ for a total of $25,000$ time steps. The interaction range is fixed at $R = 6$. To accelerate the computation, the non-local coupling term is evaluated using a two-dimensional discrete convolution:
\begin{equation*}
    (\phi * \mathcal{K})_{ij} = \sum_{i',j'} \phi_{i+i',j+j'}\, \mathcal{K}_{i',j'},
\end{equation*}
where $*$ denotes the convolution operation, and $\mathcal{K}$ is a kernel of size $(2R + 1) \times (2R + 1)$. 
To characterize the topological defects, we calculate the winding number $n$, defined as the sum of phase differences around a plaquette:
\begin{equation*}
    \sum_{i} \Delta \phi = 2\pi n,
\end{equation*}
where the summation is taken clockwise around the smallest square cell of the discrete lattice. The winding number $n$ takes integer values of $\pm 1$ or $0$.

{\it A single vortex:}
First, we investigate a single isolated vortex to validate our numerical results against the analytical solution in the small $\alpha$ limit. We employ $N = 151,\ t=100,\ \alpha = 10^\circ$ and impose reflective boundary conditions.
A $\mathit{+1}$ defect is initialized at the center of the simulation box, as shown in Fig.~\ref{fig:OP}(a) and (b).
The profiles of the amplitude $|z|$ and phase $\Phi$ along the diagonal cross-section exhibit excellent agreement with the analytical predictions shown in Fig.~\ref{fig:OP}(c) and (d).
In the coherent core region, the phase $\Phi$ remains nearly constant before dropping abruptly at $r = R$, resulting in a discontinuous derivative.
From Eq.~(\ref{eq:rho}), we obtain $\rho \approx 1.23$, which is smaller than distance $\rho_1 = \sqrt{5/2}$ 
from the vortex center to the next-nearest grid point.
This implies that the system can only form an incoherent core with just four grid points, which is 
in good agreement with the result in Fig.~\ref{fig:OP}(a).

{\it Dynamics of a vortex pair:}
In the following, we focus on vortex pairs.
Here we simulate a system size of $N = 100$ and apply periodic boundary conditions.
Because of the topological constraints of a toroidal surface, the total topological charge of the system must be zero.
To generate a vortex-antivortex pair, we construct the initial phase field using the $\arctan$ function, placing defects at coordinates $[x_1, y_1] = (N/4 + 1/2, N/4 + 1/2)$ and $[x_2, y_2] = (3N/4 - 1/2, 3N/4 - 1/2)$. 
\begin{equation*}
    \phi = \sum_{i=1,2} (-1)^{i+1}\text{arctan} \left( \frac{y - y_i}{x - x_i} \right).
\end{equation*}

We systematically investigate the spatial patterns that emerge in the system over the range of the phase lag $\alpha \in [0, \pi/2)$. Fig.~\ref{fig:2} presents the phase distribution ($\phi$), the order parameter magnitude ($|z|$), and the diagonal profile of the complex order parameter at four representative values: $\alpha = 2^\circ, 25^\circ, 45^\circ$, and $65^\circ$. 
We identify three primary dynamical regimes based on the patterns in the final state: 
(a),(b) a planar oscillation state, (c) a stable spiral wave state with incoherent cores, and 
(d) a turbulence-like state driven by the expansion of incoherent regions. 
To demarcate the parameter intervals for these distinct regimes, we define three critical values: $\alpha_0$, $\alpha_1$, 
and $\alpha_2$. These findings are in good agreement with previous studies, such as Ref.~\cite{PhysRevE.70.065201}, 
which reported similar patterns arising from random initial conditions.

As shown in Fig.~\ref{fig:2}(a), for small phase lags in the range $0 < \alpha < \alpha_0$, 
the system initially forms a pair of spiral waves. Due to the finite size of the system, 
these structures are distinct and free of incoherent cores. 
The threshold $\alpha_0$ is around $5^\circ$ but there is some ambiguity due to sporadic emergence of 
topological defects. We can estimate it from the analytical result~(\ref{eq:rho})
as $\alpha_0 \approx 8/(3 \pi R \rho_0) \approx 5.7^\circ$, where $\rho_0 = 1/\sqrt{2}$
is distance from the vortex center to the nearest grid point.
The diagonal profile of 
the order parameter confirms that $|z| \to 0$ near the center ($r = 0$) and converges to 
unity in the outer region ($r > R$), consistent with the analytical results discussed in Sec. II.

When the phase lag is in the intermediate range $\alpha_0 < \alpha < \alpha_1$ (where $\alpha_1 \approx 27^\circ$), the system develops again a pair of spiral waves [Fig.~\ref{fig:2}(b)]. However, in contrast to the previous regime, an incoherent core emerges at the center. The two vortices eventually annihilate, and the system relaxes into a planar oscillation state. We observe that the spatial variation of the order parameter becomes more gradual in this regime, indicating an increase in the size of the incoherent core. Nevertheless, for $\alpha < \alpha_1$, the system fails to sustain a stable incoherent structure in the long term.

In the regime $\alpha_1 < \alpha < \alpha_2$ (with $\alpha_2 \approx 63^\circ$), the system sustains a stable pair of spiral waves with persistent incoherent cores, as shown in Fig.~\ref{fig:2}(c). The order parameter distribution reveals an interference pattern between the spiral cores that forms a nearly square geometry, an artifact that we attribute to the square-shaped boundary conditions of the simulation. As $\alpha\approx60^\circ$, the symmetry between the two spiral waves is progressively broken, leading the system to settle into a configuration characterized by a dominant large spiral wave accompanied by one or more smaller spirals. Furthermore, during the translation of these spiral wave chimeras, distinct filamentary structures emerge within the incoherent core~\cite{bataille2023traveling}.

Finally, for large phase lags, $\alpha_2 < \alpha < \pi/2$, the system displays highly irregular patterns corresponding to a turbulent state [Fig.~\ref{fig:2}(d)]. This regime is characterized by the rapid expansion of incoherent cores, leading to the gradual degradation of coherent structures and the onset of large-scale, unpredictable dynamics. The order parameter distributions exhibit hallmark features of turbulence, including pronounced spatiotemporal fluctuations, indicating that the system has entered a highly chaotic dynamical regime.

This annihilation mechanism can be understood by considering the continuous limit. When $R=1$, the equation of motion, Eq. (\ref{eq.R}), reduces to the phase form of the Complex Ginzburg-Landau Equation (CGLE): $\frac{\partial \phi}{\partial t} = \omega - K\sin \alpha + D \nabla^2 \phi + \lambda|\nabla\phi|^2$, where $D = (K/4)\cos \alpha$ is the diffusion coefficient and $\lambda = (K/4)\sin \alpha$ governs the nonlinear effects (see Supplementary Information S3 for derivation). The Benjamin-Feir-Newell (BFN) criterion is given by $D >0$.
Consequently, when $\alpha$ is small, the system is diffusion-dominated ($D \gg \lambda$), leading to the annihilation of spirals. As $\alpha$ increases, the nonlinear term $\lambda$ becomes more significant, counteracting diffusion and allowing for the stabilization of structures.
When the coupling is non-local, although it can increase diffusion and introduce higher-order terms such as $-\nabla^4 \phi$, the most significant impact still comes from the nonlinear term, which has a fundamental effect on the system's dynamics.

\begin{figure}[h]
\includegraphics[width=0.48\textwidth]{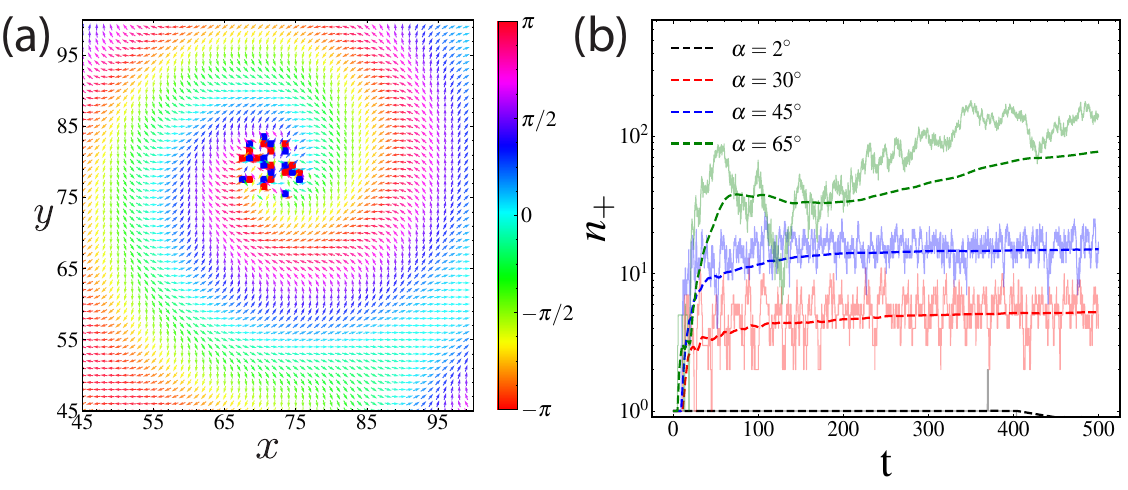}
\caption{\label{fig:3} (a) The vector field $(\cos \phi, \sin \phi)$ and topological defects in a chimera state with $\alpha = 45^\circ$, observed in the region $x,y\in [45,99]$.
(b)Time evolution of the total positive winding number $n_+$ under different values of phase lag. The y-axis is on a log scale. The dashed line shows the running time-average, $\langle n_+ \rangle (\tau) = \frac{1}{\tau} \sum_{t=1}^{\tau} n_+(t)$.
\label{fig:defects}
}
\end{figure}

The topological defects and their corresponding winding numbers are shown in Fig.~\ref{fig:3}(a). In the figure, the blue dots indicate topological defects with a winding number of $-1$, while the red dots represent those with a winding number of $+1$. 
To better understand the evolution of topological defects in the system, we select a phase lag parameter $\alpha = 45^\circ$  within the square region $x,y\in[45,99]$, 
focusing on the time evolution of an initial topological defect with winding number $-1$. To systematically analyze this characteristic, we focus on the variation of the total number of positive winding numbers $n_+$ in the system.

As shown in Fig.~\ref{fig:3}(b), we present the evolution of the total number of positive winding numbers as the coupling phase lag parameter $\alpha$ varies within the range $\alpha \in [0, \pi/2)$. We observe that in the regime before the onset of chimera states when no incoherent core has formed, the system nearly does not generate new topological defects, and the total positive winding number remains constant at its initial value of 1. However, once the system establishes a stable incoherent core , topological defects begin to be continuously generated and annihilated within this core. 
At this point, the total number of positive winding numbers dynamically fluctuates, exhibiting persistent increases and decreases within a bounded range. To further quantify the statistical features, we compute the time average of the total number of positive winding numbers, shown as the dashed lines in the figure. Within the specific parameter range $\alpha_1 < \alpha < \alpha_2$, the average number of positive winding numbers tends to saturate and fluctuates around a stable mean value. This observation suggests that the system may attain a statistical steady state in this parameter regime. In particular, even in the case $\alpha_0 < \alpha < \alpha_1$, where the two spiral waves eventually annihilate, the system still exhibits an approximately stable average winding number.
In contrast, when $\alpha > \alpha_2$, the time-averaged number of positive winding numbers continues to grow throughout the simulation and shows no sign of saturation. This behavior indicates that the system has lost its stable chimera structure and is now in a highly non-equilibrium regime, possibly exhibiting chaotic dynamics.

\begin{figure}[h!]
\includegraphics[width=0.45\textwidth]{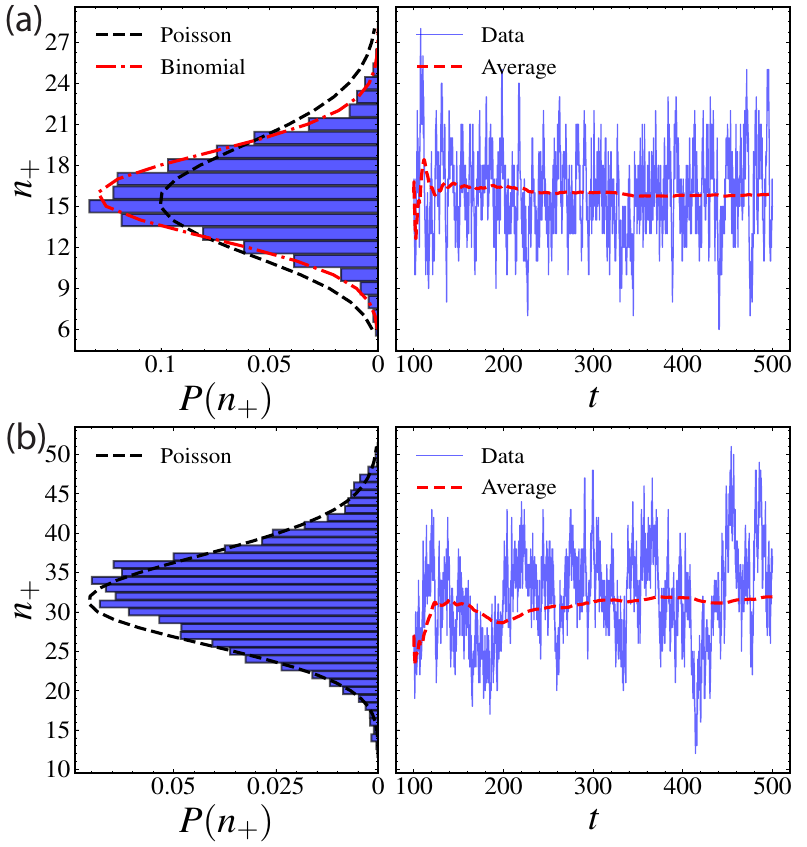}
\caption{\label{fig:4} Statistics of the total positive winding number $n_+$ under different phase lags.     
    (a) $\alpha = 45^\circ$. 
    (b) $\alpha = 55^\circ$. 
    The left panels show the histogram of $n_+$ compared to a truncated Poisson distribution [Eq.~\ref{eq.Poi}] and a binomial distribution [Eq.~\ref{eq.Binom}]. 
    The right panels show the corresponding time evolution of $n_+$ (blue line)  and its running time-average (red dashed line), $\langle n_+ \rangle (\tau) = \frac{1}{\tau} \sum_{t=100}^{\tau} n_+(t)$
    (Note that the $\alpha = 55^\circ$ case is not compared with the binomial distribution, because $\sigma^2 > \mu$).}
\end{figure}

{\it Defect Statistics.}
We conducted a statistical analysis of topological defects to elucidate their behavior under varying phase lag (see Supplementary Information S4 for details).
As shown in Fig.~\ref{fig:4}, we plot the distribution of the total positive winding number $n_+$
for the phase lag $\alpha=45^\circ$ and $\alpha=55^\circ$.
We first compare these data to the theoretical zero-truncated Poisson distribution given by:
\begin{equation}
    P(n_+) = \frac{{\mu}^{n_+} e^{-{\mu}}}{n_+! (1 - e^{-{\mu}})}, \quad n_+ \geq 1
    \label{eq.Poi}
\end{equation}
where ${\mu} \equiv \overline{n_+}$ .
This model, however, inherently assumes that the variance of the distribution is equal to its mean ($\sigma^2 = {\mu}$).
We find that our measured data exhibits a variance significantly smaller than its mean ($\sigma^2 < {\mu}$), which contradicts the Poisson assumption. This variance suppression ($\sigma^2 < {\mu}$) is the defining characteristic of the binomial distribution, suggesting a finite number of underlying sites ($N_{\text{core}}$) capable of generating defects.
Therefore, we also compare the data to a binomial distribution, $B(N_{\text{core}}, p)$, whose probability mass function is:
\begin{equation}
    P(n_+) = \binom{N_{\text{core}}}{n_+} p^{n_+} (1-p)^{N_{\text{core}}-n_+}.
    \label{eq.Binom}
\end{equation}
Crucially, the parameters $N_{\text{core}}$ and $p$ are not arbitrary fitting parameters. They are determined directly from the measured mean $\mu$ and variance $\sigma^2$ of the data segment via the relations:
\begin{equation}
    N_{\text{core}} = \frac{\mu^2}{\mu - \sigma^2} \quad \text{and} \quad p = \frac{\mu}{N_{\text{core}}}.
    \label{eq.Solve}
\end{equation}
As shown in Fig.~\ref{fig:4}(a), the binomial distribution derived from the data's own moments provides a substantially more accurate fit than the Poisson model.

\begin{figure}[h]
\includegraphics[width=0.48\textwidth]{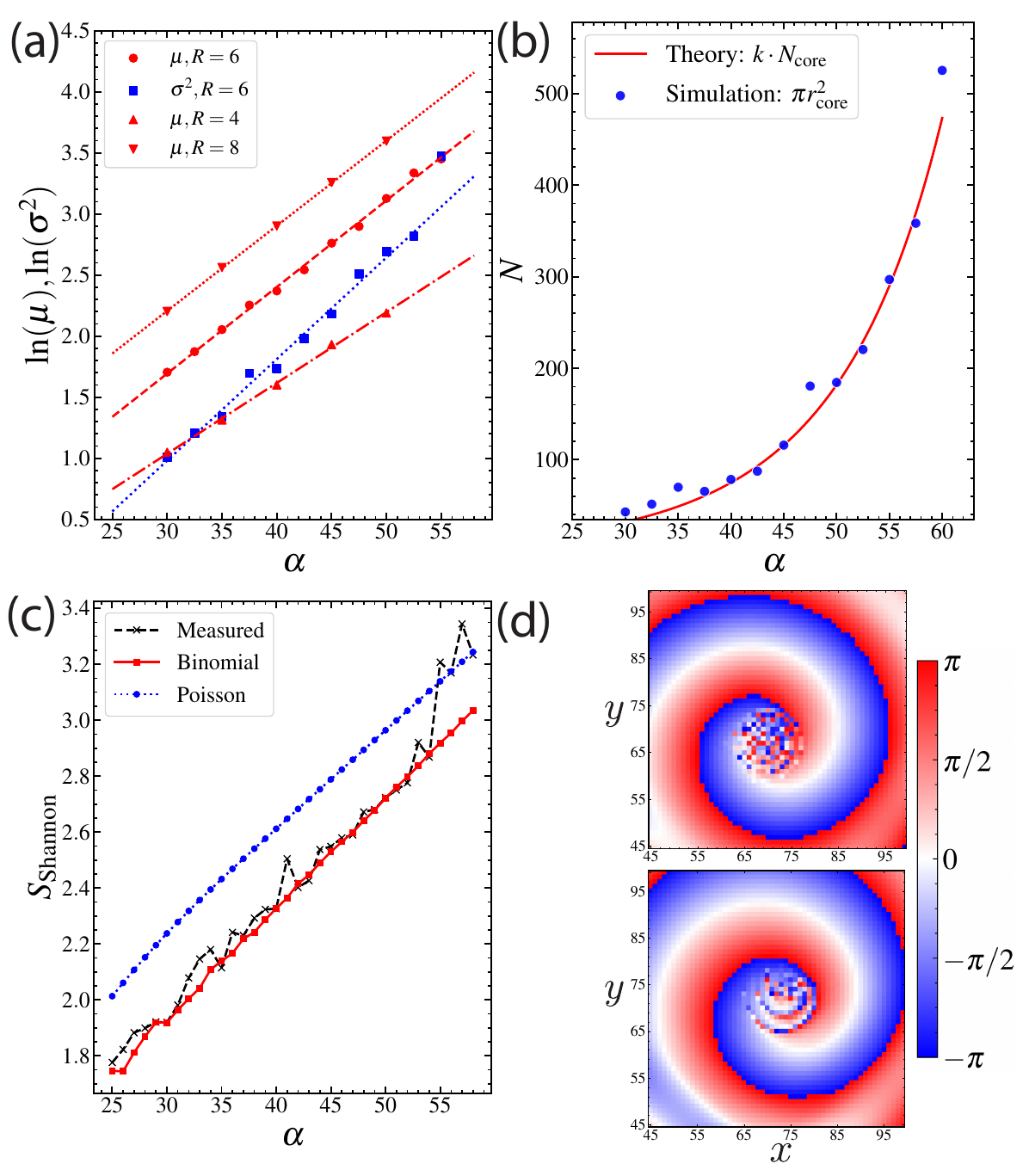}
\caption{\label{fig:5} (a) 
Dependence of the average total positive winding number $\mu$
and its standard deviation $\sigma^2$ on the phase lag $\alpha$. 
The red markers show $\mu$ for different interaction radii. 
The blue markers show $\ln(\sigma^2)$ for $R=6$. 
The lines represent linear fits to the corresponding data points.
(b) The numerical simulation results for the core area $\pi r_{\text{core}} ^2$ are compared with the theoretical prediction $N_{\text{core}}(\alpha)$ [Eq. (\ref{eq.N})]. 
(c) Comparison of the measured entropy with the theoretical binomial and Poisson entropy. 
(d) Comparison of spatial phase patterns near the critical transition point $\alpha^*$ at $t=190$, observed in the region $x,y\in [45,99]$ for $\alpha = 54^\circ$ (top) and $\alpha = 55^\circ$ (bottom).
}
\end{figure}

To quantitatively assess the influence of $\alpha$ on the system properties, we computed the corresponding $\mu, \sigma$, and the information entropy $S_{\mathrm{Shannon}} = - \sum_{n_+ = 1}^{\infty} P(n_+) \ln P(n_+)$ determined by Eq.~(\ref{eq.Poi}).
We first performed a linear fit for $\ln \mu$ and $\ln \sigma$ versus $\alpha$, as shown in Fig.~\ref{fig:5}(a). The results show strong linear correlations, with coefficients of determination exceeding 0.99. The fitting results are as follows: $\ln \mu = 12.75 \alpha/\pi - 0.43,  \ln \sigma^2 = 14.96 \alpha /\pi - 1.51,$ which implies an exponential form:
\begin{equation}
\mu = a e^{b \alpha}, \quad \sigma^2 = c e^{d \alpha},
\label{eq:fit}
\end{equation}
where $a, b, c$, and $d$ are constants determined by the fit. Substituting it into Eq.~(\ref{eq.Solve}), we get
\begin{equation}
N_{\text{core}}(\alpha) = \frac{a^2 e^{2b \alpha}}{a e^{b \alpha} - c e^{d \alpha}}.
\label{eq.N}
\end{equation}
This formula predicts super-exponential growth and divergence of the core size for increasing $\alpha$ (note that $b < d$).
In fact, $\sigma^2$ is deviated from the exponential growth law and exceeds $\mu$ at $\alpha=55^\circ$, 
where $N_{\text{core}}$ diverges according to Eq.(\ref{eq.Solve}) and the binomial statistics break down.
To verify the validity of the analytical expression for $N_{\text{core}}$, we performed numerical simulations to measure the  core area. The comparison between the numerical data and the theoretical curve is illustrated in Fig.~\ref{fig:5}(b).
 we find that the theoretical node count and the simulated area follow the scaling relationship:
$N_{\text{core}} \approx \frac{1}{k} (\pi r_{\text{core}} ^2) $ where the scaling factor $k \approx 3.02$.
This close agreement across the surveyed range of $\alpha$ demonstrates that the evolution of the spiral wave core size is quantitatively governed by the exponential competition between the statistical parameters $\mu$ and $\sigma^2$. 

To further verify the exponential relationship, we repeated the above analysis for different coupling radii $R = 4, 8$. 
We can observe that the influence of $R$ on the slope of the fitted curve is comparatively small. The intercept ($\mu$ at $\alpha=0$) increases with $R$, but the dependence is milder than the scaling of the area of the 
coupling range ($\propto R^2$); for example, the $\mu$ for $R=4$ and $R=8$ are different by 
the ratio $3.4$, which is smaller than $(8/4)^2=4$.

We also examined the effect of system size by repeating the analysis at $N = 150, 200$ and for the single-spiral case (see Supplementary Information S5 for details). 
The fitting curve remains nearly unchanged, indicating that the system size has little effect on the conclusion. 
The relationship appears to be universal.

In Fig.~\ref{fig:5}(b), we compare the measured entropy with the theoretical Poisson and Binomial entropies as a function of $\alpha$. The theoretical curves are generated by substituting the empirical fits (Eq.~(\ref{eq:fit})) into the Poisson model (Eq.~\ref{eq.Poi}) and the binomial model (Eq.~\ref{eq.Binom}), respectively.

It is observed that for $\alpha < 55^\circ$, the measured entropy agrees remarkably well with the binomial prediction. Around $\alpha \approx 55^\circ$, the system exhibits a sharp rise in entropy, reaching and even surpassing the level expected from a Poisson process. We identify this point as a new critical value, $\alpha^*$, marking a transition from binomial‐like to Poisson‐like behavior. 

To elucidate the structural origin of this transition, we compare the spatial patterns at $\alpha = 54^\circ$ and $\alpha = 55^\circ$ in Fig.~\ref{fig:5}(d). Notably, at $\alpha = 55^\circ$, the core exhibits the emergence of filamentous structures. These structural features and the identified transition point are in good agreement with previous studies on the transition from static to moving chimera states~\cite{bataille2023traveling}(see Supplementary Movie 1 for a dynamic visualization of this process). However, our study distinguishes itself in two key aspects: In previous works, the transition point is often distributed or broadened due to the inhomogeneity of natural frequencies, whereas our system of identical oscillators yields a sharper definition of the transition.
While previous coarse-grained approaches could not resolve fine-scale dynamics, our method tracks individual topological defects. Consequently, our analysis provides a clearer and more quantitative description of the static-moving chimera transition.

{\it Discussion.}
This study reveals the statistical laws governing topological defects in spiral wave chimera states and establishes the total positive winding number, $\mu$, as a robust macro-variable for characterizing these complex states. Our analysis leads to two major physical insights regarding the scaling, and the statistical transition of the chimera core.

First, we establish that in the stable chimera regime, $\mu$ follows a clear exponential growth with the phase lag ($\mu \sim e^{b\alpha}$). This relationship is concise in form and fits the data well. The phenomenon demonstrates robustness within a specific parameter range and is reproduced under various conditions, suggesting that topological excitations exhibit an underlying statistical order.
Crucially, our scaling analysis refutes the hypothesis of simple geometric growth. 
The observed divergence between defect generation and core area expansion indicates that the defect population is driven by intrinsic dynamical instability rather than passive geometric scaling.

Second, we identified a critical transition point at $\alpha^* \approx 55^\circ$, marking a fundamental shift in the nature of the core turbulence. For $\alpha < \alpha^*$, the defect statistics follow a Binomial distribution, suggesting a constrained regime where defect generation is limited by steric repulsion or core capacity. However, as $\alpha$ exceeds $\alpha^*$, we observe a sharp rise in entropy and a crossover to a Poisson-like distribution, indicating an unconstrained regime dominated by independent, random excitations. Physically, this statistical transition coincides with the emergence of filamentous structures within the core and signals the onset of the moving chimera state.

In conclusion, topological defects in chimera states are not random artifacts but possess quantifiable structural features that evolve regularly with system parameters. The transition from the geometry-dominated linear growth at small $\alpha$ to the instability-driven exponential growth at larger $\alpha$, and finally to the Poissonian structural breakdown at $\alpha^*$, delineates the rich dynamical landscape of chimera states.
A detailed investigation of how one scaling regime transitions to the other, 
near the critical values $\alpha_1$ and $\alpha_2$, remains a crucial future task.
Such a study could provide key insights into the mechanisms that stabilize and destabilize 
the spiral wave chimera states.



\bibliography{liu2025topological}

@book{Kuramoto1984,
author="Kuramoto, Yoshiki",
title="Chemical Oscillations, Waves, and Turbulence",
year="1984",
publisher="Springer Berlin, Heidelberg",
isbn="978-3-642-69689-3",
doi="10.1007/978-3-642-69689-3_7",
url="https://doi.org/10.1007/978-3-642-69689-3_7"
}

@book{strogatz2003sync,
  author    = {Steven Strogatz},
  title     = {Sync: The Emerging Science of Spontaneous Order},
  year      = {2003},
  publisher = {Hyperion},
  isbn      = {978-0-7868-6844-5},
  oclc      = {50511177}
}

@article{RevModPhys.77.137,

  author = {Acebr\'on, Juan A. and Bonilla, L. L. and P\'erez Vicente, Conrad J. and Ritort, F\'elix and Spigler, Renato},
  journal = {Rev. Mod. Phys.},
  volume = {77},
  issue = {1},
  pages = {137--185},
  numpages = {0},
  year = {2005},
  month = {Apr},
  publisher = {American Physical Society},
  doi = {10.1103/RevModPhys.77.137},
  url = {https://link.aps.org/doi/10.1103/RevModPhys.77.137}
}

@article{kuramoto2002coexistence,
  author  = {Kuramoto, Yoshiki and Battogtokh, Davaa},

  journal = {Nonlinear Phenomena in Complex Systems},
  year    = {2002},
  volume  = {5},
  number  = {4},
  pages   = {380--385},
  type    = {J}
}

@article{PhysRevE.69.036213,

  author = {Shima, Shin-ichiro and Kuramoto, Yoshiki},
  journal = {Phys. Rev. E},
  volume = {69},
  issue = {3},
  pages = {036213},
  numpages = {9},
  year = {2004},
  month = {Mar},
  publisher = {American Physical Society},
  doi = {10.1103/PhysRevE.69.036213},
  url = {https://link.aps.org/doi/10.1103/PhysRevE.69.036213}
}

@article{PhysRevLett.93.174102,

  author = {Abrams, Daniel M. and Strogatz, Steven H.},
  journal = {Phys. Rev. Lett.},
  volume = {93},
  issue = {17},
  pages = {174102},
  numpages = {4},
  year = {2004},
  month = {Oct},
  publisher = {American Physical Society},
  doi = {10.1103/PhysRevLett.93.174102},
  url = {https://link.aps.org/doi/10.1103/PhysRevLett.93.174102}
}

@article{PhysRevLett.101.084103,

  author = {Abrams, Daniel M. and Mirollo, Rennie and Strogatz, Steven H. and Wiley, Daniel A.},
  journal = {Phys. Rev. Lett.},
  volume = {101},
  issue = {8},
  pages = {084103},
  numpages = {4},
  year = {2008},
  month = {Aug},
  publisher = {American Physical Society},
  doi = {10.1103/PhysRevLett.101.084103},
  url = {https://link.aps.org/doi/10.1103/PhysRevLett.101.084103}
}

@article{PhysRevLett.110.244102,

  author = {Nkomo, Simbarashe and Tinsley, Mark R. and Showalter, Kenneth},
  journal = {Phys. Rev. Lett.},
  volume = {110},
  issue = {24},
  pages = {244102},
  numpages = {5},
  year = {2013},
  month = {Jun},
  publisher = {American Physical Society},
  doi = {10.1103/PhysRevLett.110.244102},
  url = {https://link.aps.org/doi/10.1103/PhysRevLett.110.244102}
}

@article{MAJHI2019100,

journal = {Physics of Life Reviews},
volume = {28},
pages = {100-121},
year = {2019},
issn = {1571-0645},
doi = {https://doi.org/10.1016/j.plrev.2018.09.003},
url = {https://www.sciencedirect.com/science/article/pii/S1571064518301088},
author = {Soumen Majhi and Bidesh K. Bera and Dibakar Ghosh and Matjaž Perc}
}

@article{10.1143/PTP.76.576,
    author = {Sakaguchi, Hidetsugu and Kuramoto, Yoshiki},

    journal = {Progress of Theoretical Physics},
    volume = {76},
    number = {3},
    pages = {576-581},
    year = {1986},
    month = {09},
    issn = {0033-068X},
    doi = {10.1143/PTP.76.576},
    url = {https://doi.org/10.1143/PTP.76.576},
}

@article{PhysRevE.103.032204,

  author = {Sarkar, Mrinal and Gupte, Neelima},
  journal = {Phys. Rev. E},
  volume = {103},
  issue = {3},
  pages = {032204},
  numpages = {14},
  year = {2021},
  month = {Mar},
  publisher = {American Physical Society},
  doi = {10.1103/PhysRevE.103.032204},
  url = {https://link.aps.org/doi/10.1103/PhysRevE.103.032204}
}

@article{PhysRevE.70.065201,

  author = {Kim, Pan-Jun and Ko, Tae-Wook and Jeong, Hawoong and Moon, Hie-Tae},
  journal = {Phys. Rev. E},
  volume = {70},
  issue = {6},
  pages = {065201},
  numpages = {4},
  year = {2004},
  month = {Dec},
  publisher = {American Physical Society},
  doi = {10.1103/PhysRevE.70.065201},
  url = {https://link.aps.org/doi/10.1103/PhysRevE.70.065201}
}

@article{PhysRevE.90.030902,

  author = {Rosin, David P. and Rontani, Damien and Haynes, Nicholas D. and Sch\"oll, Eckehard and Gauthier, Daniel J.},
  journal = {Phys. Rev. E},
  volume = {90},
  issue = {3},
  pages = {030902},
  numpages = {5},
  year = {2014},
  month = {Sep},
  publisher = {American Physical Society},
  doi = {10.1103/PhysRevE.90.030902},
  url = {https://link.aps.org/doi/10.1103/PhysRevE.90.030902}
}

@article{PhysRevLett.104.044101,
  author = {Martens, Erik A. and Laing, Carlo R. and Strogatz, Steven H.},
  journal = {Phys. Rev. Lett.},
  volume = {104},
  issue = {4},
  pages = {044101},
  numpages = {4},
  year = {2010},
  month = {Jan},
  publisher = {American Physical Society},
  doi = {10.1103/PhysRevLett.104.044101},
  url = {https://link.aps.org/doi/10.1103/PhysRevLett.104.044101}
}

@article{Flovik_2016,

   volume={6},
   ISSN={2045-2322},
   url={http://dx.doi.org/10.1038/srep32528},
   DOI={10.1038/srep32528},
   number={1},
   journal={Scientific Reports},
   publisher={Springer Science and Business Media LLC},
   author={Flovik, Vegard and Macià, Ferran and Wahlström, Erik},
   year={2016},
   month=sep }

@article{Totz2018,
  author  = {Totz, J. F. and Rode, J. and Tinsley, M. R. and Showalter, K. and Engel, H.},
  journal = {Nature Physics},
  year    = {2018},
  volume  = {14},
  number  = {3},
  pages   = {282--285},
  doi     = {10.1038/s41567-017-0005-8},
  issn    = {1745-2481}
}

@article{Tinsley2012,
  author  = {Tinsley, M. R. and Nkomo, S. and Showalter, K.},
  journal = {Nature Physics},
  year    = {2012},
  volume  = {8},
  number  = {9},
  pages   = {662--665},
  doi     = {10.1038/nphys2371},
  issn    = {1745-2481}
}

@article{
doi:10.1073/pnas.1302880110,
author = {Erik Andreas Martens  and Shashi Thutupalli  and Antoine Fourrière  and Oskar Hallatschek },
journal = {Proceedings of the National Academy of Sciences},
volume = {110},
number = {26},
pages = {10563-10567},
year = {2013},
doi = {10.1073/pnas.1302880110}}

@article{Kapitaniak2014,
  author  = {Kapitaniak, T. and Kuzma, P. and Wojewoda, J. and Czolczynski, K. and Maistrenko, Y.},
  journal = {Scientific Reports},
  year    = {2014},
  volume  = {4},
  number  = {1},
  pages   = {6379},
  doi     = {10.1038/srep06379},
  issn    = {2045-2322}
}

@article{PhysRevE.90.032905,

  author = {Gambuzza, Lucia Valentina and Buscarino, Arturo and Chessari, Sergio and Fortuna, Luigi and Meucci, Riccardo and Frasca, Mattia},
  journal = {Phys. Rev. E},
  volume = {90},
  issue = {3},
  pages = {032905},
  numpages = {8},
  year = {2014},
  month = {Sep},
  publisher = {American Physical Society},
  doi = {10.1103/PhysRevE.90.032905},
  url = {https://link.aps.org/doi/10.1103/PhysRevE.90.032905}
}

@article{PhysRevE.82.036202,

  author = {Lee, Tony E. and Tam, Heywood and Refael, G. and Rogers, Jeffrey L. and Cross, M. C.},
  journal = {Phys. Rev. E},
  volume = {82},
  issue = {3},
  pages = {036202},
  numpages = {10},
  year = {2010},
  month = {Sep},
  publisher = {American Physical Society},
  doi = {10.1103/PhysRevE.82.036202},
  url = {https://link.aps.org/doi/10.1103/PhysRevE.82.036202}
}

@article{PhysRevLett.127.088004,
  author = {Rouzaire, Ylann and Levis, Demian},
  journal = {Phys. Rev. Lett.},
  volume = {127},
  issue = {8},
  pages = {088004},
  numpages = {6},
  year = {2021},
  month = {Aug},
  publisher = {American Physical Society},
  doi = {10.1103/PhysRevLett.127.088004},
  url = {https://link.aps.org/doi/10.1103/PhysRevLett.127.088004}
}

@article{15cd7884-e7e8-3592-aee4-ba066c0c3e21,
 ISSN = {00361399},
 URL = {http://www.jstor.org/stable/2100638},
 author = {Donald S. Cohen and John C. Neu and Rodolfo R. Rosales},
 journal = {SIAM Journal on Applied Mathematics},
 number = {3},
 pages = {536--547},
 publisher = {Society for Industrial and Applied Mathematics},
 urldate = {2025-03-09},
 volume = {35},
 year = {1978}
}

@article{article23,
author = {Omel'chenko, Oleh and Knobloch, Edgar},
year = {2019},
month = {09},
pages = {},

volume = {21},
journal = {New Journal of Physics},
doi = {10.1088/1367-2630/ab3f6b}
}

@article{article33,
author = {Maistrenko, Yuriy and Sudakov, Oleksandr and Osiv, Oleksiy and Maistrenko, Volodymyr},
year = {2015},
month = {07},
pages = {073037},
volume = {17},
journal = {New Journal of Physics},
doi = {10.1088/1367-2630/17/7/073037}
}

@article{bataille2023traveling,
  author={Bataille-Gonzalez, M and Clerc, MG and Knobloch, E and Omel’chenko, OE},
  journal={New Journal of Physics},
  volume={25},
  number={10},
  pages={103023},
  year={2023},
  publisher={IOP Publishing},
doi = {10.1088/1367-2630/acfd4f}
}

@article{PhysRevE.94.010204,

  author = {Lau, Hon Wai and Davidsen, J\"orn},
  journal = {Phys. Rev. E},
  volume = {94},
  issue = {1},
  pages = {010204},
  numpages = {6},
  year = {2016},
  month = {Jul},
  publisher = {American Physical Society},
  doi = {10.1103/PhysRevE.94.010204},
  url = {https://link.aps.org/doi/10.1103/PhysRevE.94.010204}
}

@article{totz2018spiral,

  author={Totz, Jan Frederik and Rode, Julian and Tinsley, Mark R and Showalter, Kenneth and Engel, Harald},
  journal={Nature Physics},
  volume={14},
  number={3},
  pages={282--285},
  year={2018},
  publisher={Nature Publishing Group UK London}
}

@article{Li_2021,
ISSN={2470-0053},
url={http://dx.doi.org/10.1103/PhysRevE.104.054210},
DOI={10.1103/physreve.104.054210},
number={5},
pages = {054210},
journal={Phys. Rev. E},
volume={104},
publisher={American Physical Society (APS)},
author={Li, Bojun and Uchida, Nariya},
year={2021},
month=nov }

\end{document}